\documentclass{ws-procs975x65}
\newcommand{\be}{\begin{equation}}
\newcommand{\ee}{\end{equation}}
\newcommand{\bea}{\begin{eqnarray}}              
\newcommand{\eea}{\end{eqnarray}}

\begin{document}

\title{MG13 proceedings: 
  INFRARED ISSUES IN GRAVITON HIGGS THEORY}

\author{
  {SRIJIT BHATTACHARJEE$^{*}$; PARTHASARATHI MAJUMDAR}
}

\address{Astroparticle Physics \& Cosmology Division,\\
  ~Saha Institute of Nuclear Physics, India\\
  1/AF Bidhannagar, Kolkata-64\\
  E-mail: srijit.bhattacharjee@saha.ac.in}

\begin{abstract}
 We investigate the one-loop infrared behaviour of the effective potential in minimally coupled graviton Higgs theory in Minkowski background. The gravitational analogue of one loop Coleman Weinberg effective potential turns out to be complex, the imaginary part indicating an infrared instability. This instability is traced to a tachyonic pole in the graviton propagator for constant Higgs fields. Physical implications of this behaviour are studied. We also discuss physical differences between gauge theories coupled to Higgs fields and graviton Higgs theory.
\end{abstract}


\bodymatter

\section{Introduction}
\label{sec:intro}
The attractive nature of gravity is a source of many kind of instabilities. In classical Newtonian gravity we encounter such an instability when we treat the Universe as being filled with a static, homogeneous nonrelativistic fluid. For long-wavelength gravitational perturbations i.e. in the infrared limit the system develops an instability. This instability is very often be related to classical Jeans instability \cite{Jeans}. This is unlike in the case of an electrical plasma where charge carriers produce a screening effect over the fluid known as Debye screening. In Jeans treatment the density fluctuation satisfies the plane wave solution $\rho_1\propto \exp\{i{\bf k.x}-i\omega t\}$ which leads to a  dispersion relation, $\omega^2=v_s^2{\bf k^2}-4\pi G\rho$. Clearly, $\omega$ is imaginary for wave numbers below the critical value
\be k_J=\left(\frac{4\pi G\rho}{v_s^2}\right)^{1 \over 2}\nonumber\ee
This indicates that below this critical value we may have an exponential growth or decay of the disturbances \cite{Wein}. In the early eighties Gross et. al. \cite{Gross} showed that flat space is stable at zero temperature both classically and quantum mechanically under perturbative qantum fluctuations of Euclidean 4-space. However, when the system is kept in contact with a heat bath, the self-gravitating system becomes unstable. This kind of instability can also happen for gravitons interacting with thermally excited scalars or massless spinor fields \cite {Gross, Kikuchi} leading to a Jeans like instability. Here we show that even at zero temperature, similar instability appears in the form of a tachyonic pole in the one-loop effective graviton propagator when gravitons are coupled to a massless constant scalar background. This tachyonic pole, in turn, leads to appearance of an imaginary contribution in the one-loop effective potential for a theory where gravity is minimally coupled to a scalar. This implies that graviton fluctuations coupled to constant scalar field background
at $T=0$ in flat spacetime plays a role similar to gravitons in a finite 
temperature heat-bath inducing an instability in flat spacetime \cite{sb}.   

\section{Tachyonic mode at zero temperature propagator}
\label{sec:tachyonprop}

We now proceed to calculate the graviton propagator in one-loop approximation. The Lagrangian for Gravity coupled to a massless scalar
field,
\be\sqrt{-g}{\cal L}=\frac{1}{\kappa^2} R +\sqrt{-g}\frac{1}{2}g^{\mu
\nu}\partial_{\mu}\phi\partial_{\nu}\phi -\sqrt{-g}V(\phi)\label{Lag}\ee
We split the metric as follows,
\begin{equation}\label{expan_ep}
g_{\mu\nu}=\eta_{\mu\nu} + \kappa\, h_{\mu\nu},
\end{equation}
where the fluctuations $h_{\mu\nu}$ are small,  $|h_{\mu\nu}|<1$ and $\eta_{\mu\nu}=diag(-1,1,1,1)$.

We also expand the matter Lagrangian around a space-time constant background and retain terms upto quadratic in fluctuations.

\be \phi=\phi_0+\Phi\ee
With both the pure gravitational and matter Lagrangian expanded upto quadratic in fluctuations we can now extract the propagator for this theory easily. If we write down an effective linearized equation of motion for the graviton field from the quadratic part of the Lagrangian, we get an equation 
\be {\cal I }^{\alpha\beta,\mu\nu}h_{\mu\nu}=\kappa T^{\alpha\beta}
~, \label{eom} \ee
where $T^{\mu\nu}$ contains interaction terms containing
appropriately contracted products of terms {\it linear} in the scalar and
graviton fluctuation fields. The operator ${\cal I}^{\mu\nu,\alpha\beta}$ 
in Fourier space is given by,

\be {\cal I}^{\mu\nu\alpha\beta}=(-k^2+\kappa^2 V)\left({1\over 2} \, \eta_{\alpha\mu} \eta_{\beta\nu} 
- {1\over4}\, \eta_{\alpha\beta} \eta_{\mu\nu}\right)\label{operator}\ee

Now we can invert this operator to get the propagator,
\be
D_{\mu\nu\alpha\beta}(k) =
{ \eta_{\mu\alpha} \eta_{\nu\beta} +
\eta_{\mu\beta} \eta_{\nu\alpha} - \eta_{\mu\nu} \, \eta_{\alpha\beta} 
\over k^2 -\kappa^2 V(\phi_0)}
\label{propagator}
\ee
where we have used the following gauge fixing Lagrangian,
\be {\cal
L}_{gf}=-\frac{1}{2}\left[\partial_{\mu}h^{\mu\nu}-\frac{1}{2}\partial^{
\nu}h\right]^2
\ee
This propagator clearly contains tachyonic poles in the infrared limit for positive $V$ and can be easily  determined from the dispersion relation $k_0=\pm\sqrt{{\bf k}^2-\kappa^2 V}$. 

This tachyonic mode in the propagator is reminiscent of the Jeans Like instability that occurs in the case of gravitons interacting thermally with massless bosons or fermions.

\section{One-loop Effective potential for graviton-Higgs theory}
\label{sec:oneloop-ep}
In this section we will see that the instability in the graviton propagator also shows
up in the quantum effective potential with the appearance of an imaginary part. The starting point of computation is the Euclidean path integral 

\be Z=\int {\cal D}g {\cal D} \Phi e^{-S_E}=e^{-W}\label{PI}
\ee
The one-loop effective potential is computed employing the loop expansion scheme \cite{Jackiw} and is given by 

\small{\begin{eqnarray}
 V_{eff}^1(\phi_0)= &V&\nonumber+\frac{9}{32\pi^2}
\left[\frac{\kappa^4V^2}{2}\left(\ln{\frac{
\kappa^2V}{\Lambda^2}}-\frac{1}{2}\right)-
\kappa^2V\Lambda^2\right]-\frac{i 9\kappa^4V^2}{64\pi}\\
&+&\nonumber\frac{1}{32\pi^2}\left[(V^{''}-\kappa^2V)
\Lambda^2+\frac{a^2-2b}{4}\left(\ln{\frac{b}{\Lambda^4}}-1\right)\right]
\\&+&\frac{a\sqrt{a^2-2b}}{64\pi^2}\ln{\left[\frac{a+\sqrt{a^2-4b}}{a-\sqrt{
a^2-4b}}\right]} ~\label{effpot}
\end{eqnarray}} 
where

 {\begin{eqnarray*}a&=&V^{''}-\kappa^2 V \\
b&=&\kappa^2(2V^{' 2}-V V^{''})
\end{eqnarray*}}

The source of the imaginary part in the effective potential is the infrared sector of the functional integrals. The functional traces become non-analytic in the infrared limit. Another interesting aspect of the graviton-Higgs theory is the quantum corrections in the effective potential go away if we don't include any potential term for the Higgs field. This is true only for constant scalar background. However, in standard electroweak theory in flat spacetime, in contrast, the classical Lagrangian has Higgs-gauge field
seagull terms which lead to the one loop effective potential
\cite{sb-pm} even in the absence of a classical potential. 

In electroweak theory the existence of a non-vanishing constant Higgs field background itself is known to signify a vacuum instability which the theory resolves itself by producing masses for the gauge bosons via the Higgs mechanism. The additional instability  in this case may have originated from that vacuum instability itself, although it is obvious that this does not resolve itself by generating a mass for the gravitons. We don't have a viable picture yet to relate this instability to the structure formation of Universe via Jeans instability.

\end{document}